# Soft X-ray high-harmonic generation in an anti-resonant hollow core fiber driven by a 3 µm ultrafast laser

D. Morrill[1,†], W. Hettel[1,†,*], D. Carlson[1,†], B. Shearer[1], C. Klein[1], J. Thurston[1], G. Golba[1], R. Larsen[1], G. Seifert[1], J. Uhrich[1], D. Lesko[2,3], T. N. Nguyen[3], G. Arisholm[4], J. Knight[5], S. Diddams[2,3], M. Murnane[1], H. Kapteyn[1,6] and M. Hemmer[1]

[1]*JILA & STROBE NSF Science & Technology Center, University of Colorado & NIST, Boulder, CO 80309, USA*
[2]*Electrical, Computer and Energy Engineering, University of Colorado, Boulder, CO 80309, USA*
[3]*Department of Physics, University of Colorado, Boulder, CO 80309, USA*
[4]*FFI (Norwegian Defence Research Establishment), Postboks 25, NO-2027 Kjeller, Norway*
[5]*Centre for Photonics and Photonic Materials, Department of Physics, University of Bath, Bath BA2 7AY, UK*
[6]*Kapteyn-Murnane Laboratories, Inc. 4775 Walnut Street, #102, Boulder, CO 80301, USA*
† *These authors contributed equally to this work*
*<u>will.hettel@colorado.edu</u>

**Abstract:** High-harmonic upconversion driven by a mid-infrared femtosecond laser can generate coherent soft X-ray beams in a tabletop-scale setup. Here, we report on a compact ytterbium-pumped optical parametric chirped pulse amplifier (OPCPA) laser system seeded by an all-fiber front-end and employing periodically-poled lithium niobate (PPLN) nonlinear media operated near the pulse fluence limits of current commercially available PPLN crystals. The OPCPA delivers 3 µm wavelength pulses with 775 µJ energy at 1 kHz repetition rate, with transform-limited 120 fs pulse duration, diffraction-limited beam quality, and ultrahigh 0.33% rms energy stability over >18 hours. Using this laser, we generate soft X-ray high harmonics (HHG) in argon gas by focusing into a low-loss, high-pressure gas-filled anti-resonant hollow core fiber (ARHCF), generating coherent light at photon energies up to the argon L-edge (250 eV) and carbon K-edge (284 eV), with high beam quality and ~1% rms energy stability. This work demonstrates soft X-ray HHG in a high-efficiency guided-wave phase matched geometry, overcoming the high losses inherent to mid-IR propagation in unstructured waveguides, or the short interaction lengths of gas cells or jets. The ARHCF can operate long term without damage, and with the repetition rate, stability and robustness required for demanding applications in spectro-microscopy and imaging. Finally, we discuss routes for maximizing the soft X-ray HHG flux by driving He at higher laser intensities using either 1.5 µm or 3 µm—the signal and idler wavelengths of the laser.

## 1. Introduction

Recent advances in the generation of coherent short wavelength pulses using high-harmonic generation (HHG), high-brightness synchrotrons and X-ray free electron lasers make it possible to probe the structure and function of complex material systems [1–4], providing rich new information not accessible using conventional spectro-microscopies. In recent decades, high-harmonic generation (HHG) technology driven by near-infrared (IR) lasers has matured considerably, resulting in widespread availability of extreme-UV (EUV) light in laboratory settings. Tabletop HHG sources of EUV light have been used to demonstrate powerful new capabilities, including diffraction-limited coherent imaging at short wavelengths, ultraprecise metrologies of industrially-relevant semiconductors and nanostructured metamaterials, as well as new understanding of spin, charge and heat transport in nanostructured materials [1,2,4–13]. The EUV spectral region has proven ideal for probing samples in reflection, since materials exhibit significant reflectivity at wavelengths ~10-50 nm (corresponding to photon energies of ~20-100 eV). Many of these applications experiments require long-term stable or noise-cancelled HHG sources with ~1% rms stability over many hours.

Meanwhile, the development of soft X-ray (SXR) high-harmonic sources with the robustness, stability and flux to support applications has remained a technical challenge, because HHG scaling requires the use of intense mid-IR laser pulses. SXR harmonics will enable higher spatial resolution and penetration power for transmission microscopies compared with EUV HHG, as well as enhanced spectro-microscopies to probe core-level resonances that encode the chemical and magnetic state of each element in a material. A practical SXR HHG source requires a mid-infrared (IR) drive laser operating at high repetition rate (≥kHz), and capable of generating a focused intensity sufficient to field-ionize a high pressure (multi-atmospheres) noble gas over an extended ~5-10cm length [14–16]. This ensures phase-matched buildup of the coherent SXR signal, and helps to compensate for the decreased single-atom HHG yield (~$\lambda^{-6}$) for SXR HHG, which is caused by increased quantum diffusion of the ionized electron wave packet during its "boomerang" excursion that creates HHG in a recollisional process.

Several pioneering high-harmonic sources have generated SXR beams at near-kHz repetition rates [14–26]. For example, Ti:sapphire laser driven optical parametric amplifiers (OPA) have generated SXR HHG radiation in an extended waveguide geometry with photon flux up to $10^9$ ph/s/1% bandwidth, enabling high quality X-ray-absorption spectroscopy in the water window (284–541 eV) [17]. The same approach using a gas jet geometry achieved a photon flux of $7\times10^7$ ph/s [18,19]. However, for all of these implementations, the complexity of these multistage laser systems, as well as challenges in scaling fs-pumped OPAs to high pulse energies and repetition rates, precludes widespread application. Several other laser architectures have driven SXR HHG at wavelengths ≥~2 µm, utilizing gas-filled waveguide, gas cell and gas jet interaction geometries. Direct amplification of 2.5 µm wavelength, 7 mJ energy pulses at 1 kHz in bulk chromium-doped chalcogenide gain media in combination with a pulsed neon gas jet has resulted in HHG with ~$7\times10^4$ ph/s photon flux at 300-600 eV [20]. In another approach, 1.9 µm wavelength, 290 µJ energy pulses at 98 kHz from a thulium-fiber amplifier were coupled into an anti-resonant hollow core fiber (ARHCF) where the pulses underwent self-compression and generated high harmonics at the end of the fiber with an estimated flux of $2.8\times10^6$ ph/s/eV at 300 eV [21]. Finally, we note that three relevant optical parametric chirped pulse amplifier (OPCPA) systems pumped at 1 µm wavelength and operating slightly above 2 µm wavelength have also been explored: a 100 kHz system generated harmonics up to 600 eV in a 1.2 mm, helium filled gas cell with a total photon flux of ~$6\times10^6$ ph/s [22]; a 10 kHz system generated harmonics up to 550 eV in a 6 mm, helium-filled gas cell with an estimated flux of ~$(2-5)\times10^4$ photons/sec/eV at 284 eV [23]; and a 1 kHz system generated harmonics up to ~450 eV in a neon filled semi-infinite gas cell with an estimated flux of $1.5\times10^6$ photons/sec/1% bandwidth at 350 eV [24]. It must also be noted that SXR flux estimates are difficult and often inaccurate in this spectral region, and that many of these results have not been proven in long-term operation.

In this paper, we present a long-term ultrastable 3 µm wavelength OPCPA that delivers diffraction-limited and transform-limited optical pulses with 775 µJ energy, 120 fs duration at 1 kHz repetition rate, and with ultrahigh 0.33% rms energy stability measured over >18 hours and hands-free operation for many months. Leveraging the exceptional, near turn-key stability of this system, we present the first demonstration of mid-infrared ($\lambda \geq 3$ µm, ISO 20473) driven SXR HHG beams produced from a low-loss gas-filled anti-resonant hollow core fiber (ARHCF). The ARHCF overcomes the high losses inherent to mid-IR propagation in unstructured waveguides, or the short interaction lengths of gas cells or jets, and performed for months of regular use without optical damage. Robust SXR HHG is demonstrated up to the argon L-edge at 250 eV using argon, and up to 280 eV in nitrogen, with high beam quality and 1.1% rms energy stability – representing the first demonstration of mid-IR driven HHG at kHz repetition rates. This work demonstrates soft X-ray high-harmonic generation in a high-efficiency guided-wave phase matched geometry, and with the high repetition rate, stability and robustness required for demanding applications in spectro-microscopy and imaging. Finally, we discuss routes for achieving the maximum theoretical soft X-ray high-harmonic flux in a He-filled ARHCF driven at higher laser intensities, either by the signal beam at 1.5 µm, or by the idler at 3 µm.

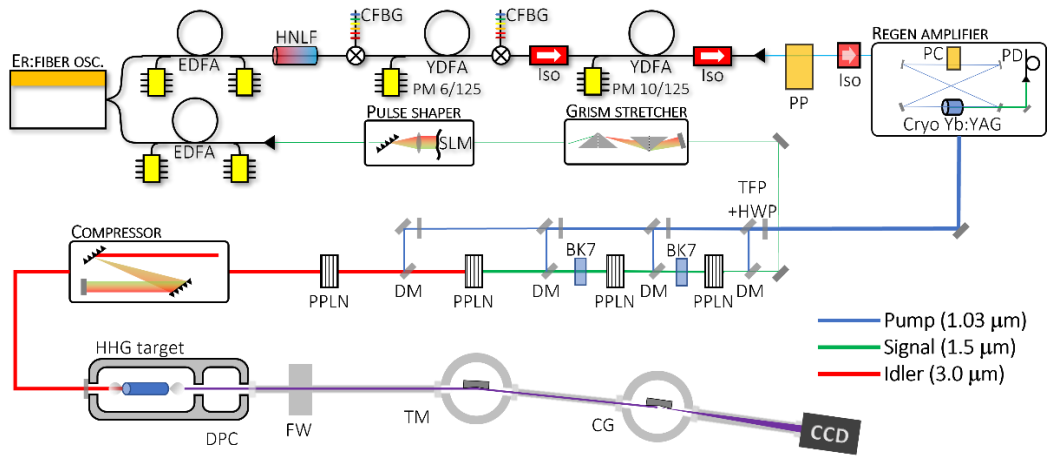

Fig. 1. Compact OPCPA system and SXR beam line including the fiber seed laser, the cryogenic Yb:YAG regenerative amplifier pump laser, the OPA chain, along with the pulse stretcher, shaper and compressor. EDFA: Er-doped fiber amplifier, YDFA: Yb-doped fiber amplifier, Iso: fiber or free space isolator, CFBG: chirped fiber Bragg grating, PP: pulse picker, PC: Pockels cell, PD: pump diode, SLM: spatial light modulator, TFP + HWP: thin film polarizer + half wave plate, BK7: idler absorber, PPLN: 5% magnesium oxide-doped periodically poled lithium niobate, DM:

dichroic mirror, DPC: differential pumping chamber, FW: filter wheel, TM: toroidal mirror, CG: curved grating. The SXR beam line consists of an anti-resonant waveguide chamber, followed by a filter wheel to reject the laser light, a toroidal mirror, a curved grating and a CCD camera.

## 2. Mid-IR OPCPA architecture and performance

The OPCPA architecture consists of a fiber laser front-end, followed by a four-stage optical parametric amplifier (OPA) system pumped by a cryogenically cooled Yb:YAG regenerative amplifier (Fig. 1). In contrast with many OPCPA systems described in the literature where the chirped pulses are relatively short and that consequently require relatively large aperture amplifiers, in this system the ~0.3 ns pulse duration allows the use of relatively small beams in a collinear geometry that allows for near diffraction limited beam quality. Furthermore, the OPCPA system design completely avoids any long-path-length optical delays, stretchers, or compressor, making for a compact, practical optical layout with exceptional stability.

The front-end fiber laser is seeded by a commercial modelocked Er-doped fiber oscillator (ELMO – Menlo Systems GmbH) [27] with two outputs delivering optical pulses with ~30 pJ energy at 100 MHz and a 40 nm-FWHM spectrum centered at 1570 nm (Fig. 2 – bottom, shaded region). The pulses are chirped with ~700 fs duration. The first output of the oscillator provides the signal seed pulses for the OPCPA. Pulses from the oscillator are injected into a home-built dispersion-managed Er-doped fiber amplifier (EDFA) and simultaneously amplified, spectrally broadened, and temporally compressed [27]. The EDFA is core pumped in the forward and backward direction by two 950 mW, 976 nm wavelength diodes and uses polarization-maintaining, 4 μm core diameter, 125 μm cladding diameter gain fiber with normal dispersion. Upon exiting this EDFA into free-space, the amplified pulses are 2 nJ, with a spectrum spanning >150 nm (Fig. 2 – bottom), and near transform-limited 35 fs pulse duration. Critically, the output power stability of this amplifier was measured to be 0.2% rms over 72 hours (Fig. 2 – top) and has operated hands-off with an uptime of ~97% over the past 4 years.

The amplified 1.5 μm wavelength pulses are then directed to a home-built, spatial light modulator (SLM)-based pulse shaper. The spectral phase imparted by the SLM is controlled via software that can optimize the second-harmonic signal of the amplified 3 μm *idler* output of the OPCPA, providing near transform-limited duration of the laser output. Note that the spectral phase relationship between signal and idler wave in an OPA pumped by a narrowband pump allows us to optimize the compression of the 3 μm output by using mature 1.5 μm wavelength SLM technology.

Upon exiting the pulse shaper, the signal seed pulses are stretched in time using a grism setup that provides negative group delay dispersion (GDD) and third-order dispersion (TOD) and consists of four prisms and two gratings, arranged as shown in Fig.1. This setup provides the GDD and TOD of appropriate sign and magnitude to facilitate compression of the 3 μm idler in a Treacy-type compressor. The throughputs of the pulse shaper and grism stretcher are both ~50%. The amplified, spectrally broadened, temporally shaped and stretched 1.5 μm wavelength pulses are then directed to the OPA chain.

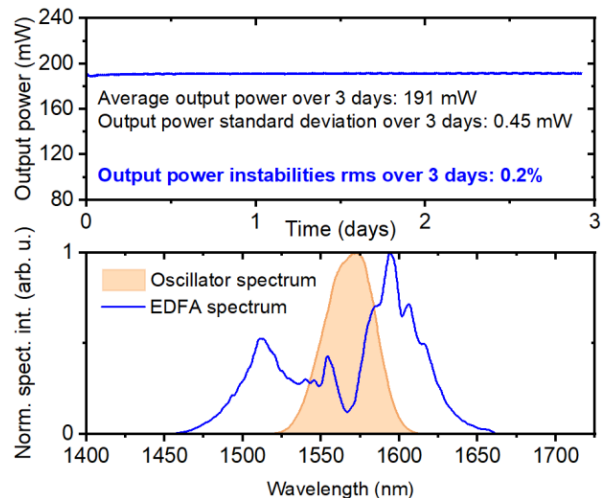

Fig. 2 (Top) Measured output power stability of the EDFA providing the signal seed pulses for the OPCPA chain showing rms stability of 0.2% over 3 days; (bottom) typical measured spectrum used to seed the OPCPA chain (blue line) along with the spectrum delivered by the commercial oscillator (shaded orange).

In parallel with this seed pulse conditioning, the second arm of the Er-fiber oscillator is used to seed the *pump* for the OPCPA. Pulses are sent into an EDFA identical to the one used for the signal beam to produce 35 fs, 2 nJ pulses. These pulses are sent into a short polarization-maintaining highly nonlinear fiber (HNLF) to generate a dispersive wave with a spectral peak centered at 1.03 μm wavelength. The optimum length of the fiber (~3.5 cm) was calculated using the pyNLO computation package that simulates nonlinear propagation in fibers accounting for $\chi^{(2)}$ and $\chi^{(3)}$ effects as well as linear dispersion [28]. The 1.03 μm dispersive wave, spanning ~90 nm spectral width, is sent into a chirped fiber Bragg grating that stretches the pulses to ~915 ps duration and narrows the spectrum to 1.2 nm bandwidth. These stretched pulses are injected into a single mode, polarization-maintaining Yb-doped fiber amplifier (YDFA, PM 6/125) pumped by a single mode, 976 nm wavelength, 950 mW diode. The stretched pulses are amplified to 270 mW average power at 100 MHz and directed to a second CFBG – identical to the first one – before entering a second and final YDFA. This second YDFA features a 10 W multi-mode pump diode that pumps a PM 10/125 double clad gain fiber. The output of this amplifier is released into free-space and can deliver up to 6 W average power. Most importantly, the power stability of this amplifier has been measured at <0.08% rms over 60 hours and has achieved >95% hands-off uptime over the past 4 years. The fiber oscillator and amplifiers all fit into a compact 30 cm by 45 cm footprint and run on battery power, making the entire front-end robust to power disruptions. After exiting the second YDFA, the amplified, stretched and $TEM_{00}$ pulse train is pulse picked from 100 MHz to 1 kHz repetition rate using an electro-optic pulse picker. 20 nJ pulses are sufficient to seed our Yb:YAG regenerative amplifier while avoiding pulse bifurcation [29].

The choice of cryogenically cooled Yb:YAG for the high-energy regenerative amplifier was guided by its excellent thermo-optic properties, the 4-level nature of the gain medium at cryogenic temperature, and the low (<9%) quantum defect, all supporting high average power applications. The regenerative amplifier cavity consists of a 4-mirror ring containing two thin film polarizers and a BBO Pockels cell operated in half-wave voltage. The pump power is provided by a 940 nm wavelength, 200 W diode array. A collimator and a relay telescope deliver the beam into the crystal with a beam diameter of 2 mm in a flat-top profile. The gain medium is a 10%-doped, 5 mm-long Yb:YAG crystal mounted in a closed-loop pulse-tube cryogenic cooler. The regenerative amplifier is pumped with 50 W of optical power and the pulses travel 20 times through the gain medium resulting in 20 mJ optical pulses with ~275 ps duration, 0.3 nm bandwidth at 1029.3 nm wavelength and a $TEM_{00}$ spatial intensity profile at 1 kHz repetition rate. The power stability at the output of the amplifier was measured to be <0.2% over 50 hours (Fig. 3 – top), the pulse-to-pulse stability <0.5% over 10 minutes and the pointing stability – recorded in the focal plane of a 50 cm focal length lens (Fig. 3 – right inset) – to be <3.5 μrad over 50 hours, representing 1.1% of the beam size at focus (Fig. 3 – middle and bottom). This amplifier has been operating in our laboratory over the past 4 years and runs hands-off, 24/7 for ~6 weeks at a time, limited by the slow accrual of water in the crystal chamber. Thermally cycling the chamber under turbo-pumping clears the contaminant.

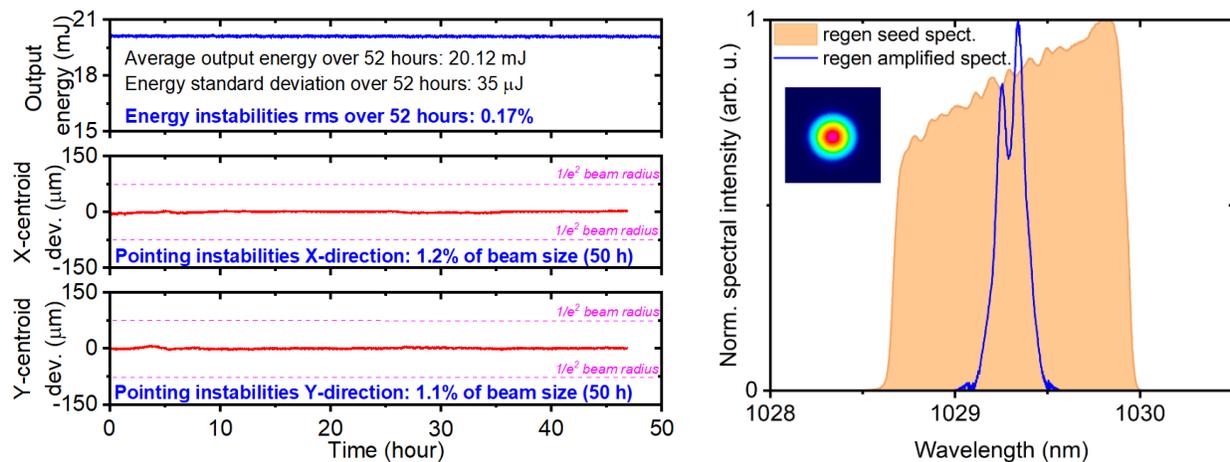

Fig. 3. (Top left) Measured energy at the output of our Yb:YAG cryogenic amplifier showing an energy stability <0.2% rms over 50 hours at the 20 mJ level; (left middle & bottom) measured pointing stability at the focus of 50 cm focal length lens showing rms deviation of ~1.15% (1.72 μm) – averaged over both axis – of the beam size over 50 hours; (right) measured seed spectrum injected into the regenerative amplifier (orange area) and measured amplified spectrum (blue line); (inset) measured far-field spatial intensity profile of the regenerative amplifier.

The OPCPA chain consists of four consecutive 5% magnesium oxide-doped periodically poled lithium niobate (MgO:PPLN) crystals. PPLN was chosen for its high nonlinearity that allows for OPCPA pumping directly by the 275 ps pulses from the regen, avoiding the need for a bulky compressor for this (narrowband) pump beam. The first stage is a 5 mm long, 29.8 μm period MgO:PPLN crystal that is tilted off-normal to suppress amplification of internally reflected pulses [30]. The first OPA is pumped by 0.2 mJ energy at a fluence of 1.4 J/cm$^2$, providing a gain of ~28000 and yielding ~10 μJ of energy in the signal wave with a spectral bandwidth of ~70 nm. These pulses are relay imaged into the second OPA stage. Dichroic mirrors are used to couple and separate the pump and signal beams at each OPA stage, and the weak idler wave is absorbed in a BK7 window – antireflection coated for the signal wave – placed after the first and second OPA stages. The second and third OPA stages each use 3 mm long, 30.1 μm period MgO:PPLN crystals pumped with 1.2 mJ and 4 mJ and feature a gain of 16 and 5 respectively. The third OPA stage is operated in the depletion regime with 20% of the pump energy being transferred to the signal beam, yielding >800 μJ pulses at 1.5 μm with ~70 nm of bandwidth. After the third OPA stage, the signal wave is discarded and the idler is used to seed the final OPA stage. This final stage consists of a 2 mm long, 3 mm aperture, 30.1 μm period MgO:PPLN and is pumped at a fluence of 1 J/cm$^2$ with up to 10 mJ energy. This final stage was thoroughly modeled using the Sisyfos code package [31] to ensure optimum extraction efficiency and bandwidth, which also resulted in reduced parasitic photorefraction effects. This optimization allows the extraction of up to 1.3 mJ energy at 3 μm wavelength with >200 nm spectral bandwidth (10% extraction efficiency from the pump to idler wave in the final stage) with an excellent beam profile and without damage to the crystal.

Upon amplification, the 3 μm idler wave is directed to a Treacy-type compressor consisting of two silver coated 300 l/mm gratings. The efficiency of the compressor is 64%, limited by the few percent loss on each silver coated optic and the diffraction efficiency of the gratings in the compressor. The beam emerging from the compressor was thoroughly diagnosed in the temporal domain using a home-built second-harmonic generation frequency resolved optical gating (SHG-FROG) apparatus, in the spatial domain using a microbolometer detector, and in output energy using a thermal power meter. The compressed output of the system reaches up to 775 μJ energy with measured power instabilities <0.33% over 18 hours and can be maintained 24/7 for weeks, completely hands-off. The spatial profile after compression shows a near-diffraction limited beam with a measured M$^2$ of 1.06 and 1.07 along the X and Y direction respectively. The rms pointing stability was measured to be <0.34% (X) x 0.43% (Y) of the beam diameter at the focus of a 50 cm focal length lens over 24 hours. Finally, in the temporal domain, after optimization with the pulse shaper, we measured 120 fs duration pulses reaching the transform limit (Fig. 4).

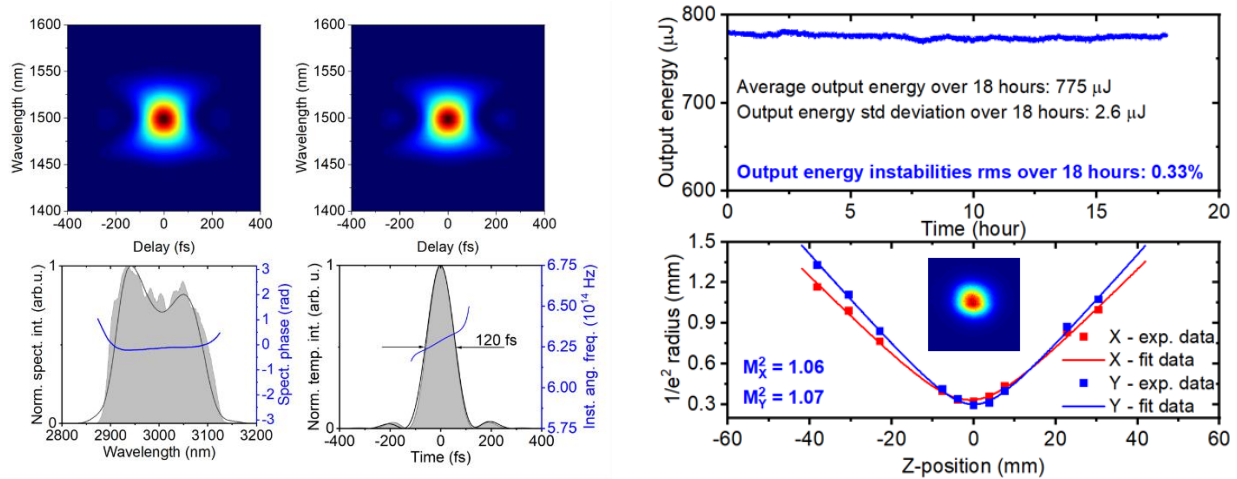

Fig. 4 (Left) Measured (top-left) and retrieved (top-right) SHG FROG traces recorded at full power along with (bottom-left) measured (shaded region) and retrieved spectral intensity (black line) and phase (blue line) showing a ~200 nm spectrum centered at 3 μm wavelength and near perfect agreement between the measured and retrieved spectra; (bottom right) retrieved temporal intensity profile from the SHG-FROG measurement (black line) overlaid with the computed Fourier transform of the experimentally measured spectrum (shaded region) along with instantaneous frequency (blue line); (right) (top) measured average output power showing power stability of 0.33% rms over 18 hours at 775 μJ output energy; (bottom) measured M$^2$ showing near diffraction-limited beam quality at 3 μm wavelength with (inset) measured spatial intensity profile in the far-field.

## 3. Ultrastable high-harmonic generation in an anti-resonant hollow core fiber

The sub-mJ, 12-cycle, 3 µm pulses at 1 kHz repetition rate are focused into a high-pressure, gas-filled, anti-resonant hollow core fiber (ARHCF) to produce phase-matched high harmonics up to the carbon K-edge at 284 eV. The mounting of the ARHCF is engineered to handle high gas pressures (up to ~35 bar) required for bright SXR HHG [14]. The ARHCF is a key enabling technology for mid-IR driven HHG, enabling a centimeter-scale interaction length using ≲ mJ pulse energy. In comparison, using an unstructured hollow core fiber (i.e. a capillary waveguide) would result in high propagation loss [32]. After the ARHCF, the HHG beam passes through two differential pumping apertures to minimize residual gas absorption. The HHG beam then passes through a filter wheel that can insert up to four filters at a time, to block both residual IR and low-order (perturbative) harmonic light and to insert samples for spectroscopy and spectral calibration. A $SiO_2$ filter is used to distinguish between HHG and low-order harmonic generation, which can penetrate very thin metallic filters. The HHG beam then reflects from a 50 cm focal length, nickel-coated toroidal mirror at ~5.3° grazing angle of incidence to relay the beam into a home-built spectrometer. The spectrometer is composed of a curved SXR grating (Hitachi 001-0437, 1200 l/mm) and an Andor Newton 940 CCD camera (Fig. 1).

The ARHCF used (iXblue IXF-ARF-120-400) supports a 90 µm mode-field diameter, is cleaved to 13 mm length with a ~45 µm diameter gas inlet hole laser-drilled into the side ~5 mm from the output. The HHG spectra at the optimal gas backing pressure of 10 bar extends to the L-edge in argon, and to the predicted phase-matching cutoff energy near the carbon K-edge in $N_2$ (Fig. 5) [9-11]. As shown in Fig. 5, the observed argon HHG spectrum is in excellent agreement with our simulations (detailed in the next section). The spectral intensity is adjusted to account for measured filter transmissions as well as modeled and manufacturer-provided efficiencies of the toroidal mirror and curved grating. The stability of the harmonic spectral intensity (Fig. 5) is measured to be 1.1% rms over the span of one hour with 10 s exposures. This overall stability is comparable to the EUV HHG stability that has successfully been used for many demanding applications in spectroscopy, metrology and imaging. The robustness of the mid-IR driving laser and careful design of the ARHCF HHG target enable the generation of HHG beams over week-long periods, hands-off. Minimal degradation of the ARHCF target was observed over the course of several months of operation.

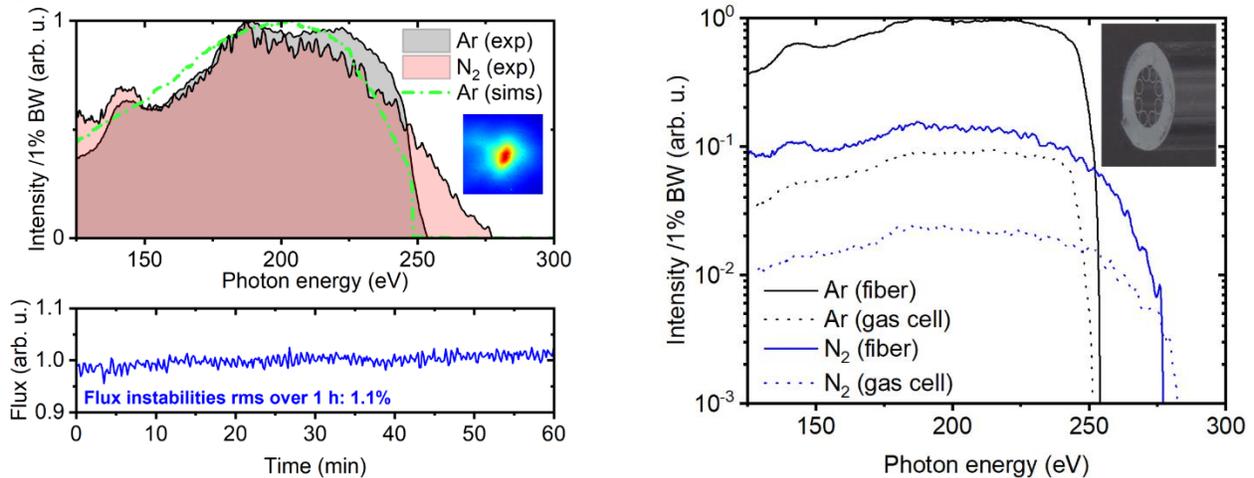

Fig. 5. High-harmonic spectra and stability. (Top left) Measured high-harmonic spectra from an anti-resonant hollow core fiber in argon and nitrogen gas, along with simulated HHG emission from argon; (inset) HHG beam from argon; (bottom left) stability of high-harmonic spectrum from argon in the ARHCF, integrated from 125 to 250 eV using 10 s exposures and showing a standard deviation of 1.1%; (right) measured spectra comparing anti-resonant hollow core fiber and gas cell HHG yield in both argon and nitrogen with (inset) image of ARHCF input.

The peak HHG flux in Ar is estimated to be ~$10^5$ photons/s/1% bandwidth, limited by the high SXR absorption of Ar gas. By switching to He, which is far less absorbing than Ar [14,16], the HHG flux is predicted to increase by ~$10^4$, but would require several times more laser intensity – for example by increasing the pulse energy to ~2 mJ and decreasing the pulse duration to the optimal ~6 cycles. However, the current 3 µm idler pulse energy is capped at 775 µJ, which is the maximum that we can extract from commercially available 3 mm aperture MgO:PPLN crystals (used here without damage). We also note that the fiber used here exhibits the smallest mode field diameter (90 µm) of any commercially available 3 µm ARHCF [17].

Rapid gas target optimization and reliable flux comparison between interaction geometries was enabled by a modular, precision-machined HHG source that can accept a wide variety of targets. With minimal modifications to the experiment, we also generated SXR HHG beams in an argon-filled gas cell with 150 µm diameter apertures, a length of 1.75 mm, and at a lower optimal backing pressure of ~2.3 bar. The observed HHG flux was more than an order of magnitude lower than in the ARHCF (Fig. 5 – right, dotted line), due to the shorter interaction length, lower gas pressure, and higher re-absorption of the generated SXR light. This geometry is not generally scalable to longer interaction lengths in He. In both geometries, the flux in $N_2$ is ~5 times lower than in Ar, likely due to molecular alignment effects [33].

### 4. Scaling soft X-ray high-harmonic generation in ARHCFs

To guide understanding and optimization of SXR HHG, we developed simulations that calculate the phase-matched HHG yield while incorporating non-adiabatic and nonlinear propagation of the driving laser field. We utilize the radially symmetric Nonlinear Envelope Equation, or (2+1)D NEE, to efficiently propagate the fundamental field [34,35]. The assumption of radial symmetry is a good approximation even for asymmetric anti-resonant fibers because of the high symmetry of the fundamental mode, allowing orders of magnitude faster runtime than using the (3+1)D NEE.

In our simulations, linear propagation is applied to the modes of the fiber, which are assumed to be the same as for unstructured capillary of the same effective radius. Since the pulse bandwidth is within the anti-resonant bands, the dispersion of the ARHCF is equal to that of a capillary [36], which is calculated with the Marcatili-Schmeltzer model [37]. Absorption is set to be zero for each mode, in agreement with the fiber's specifications for few-cm propagation. Nonlinear terms include the optical Kerr effect, plasma defocusing/absorption, and ionization loss, and are applied in the spatial domain with a split-step method. Ionization rates are calculated using the Ammosov-Delone-Krainov (ADK) equation [38] with barrier-suppression correction [39].

After we calculate the electric field at each point in space and time $E(r,t,z)$ using the field envelope and carrier-envelope phase, we use an analytical multiple pulse interference model based on the semi-classical HHG model to calculate the harmonic fields $E_h(r,\omega,z)$ [40,41]. Propagating these fields to the far field gives the full HHG spectrum. The excellent agreement between our simulations and the measured HHG spectra shown in Fig. 5, as well as through other comparisons with published data [14], was used to check the accuracy of the models.

In Fig. 6, we plot simulated HHG spectra in He driven by the current signal (1.5 µm) and idler (3.0 µm) beams i.e. using pulse parameters matching the current experiment, for different structured and unstructured fibers. Using a 64 µm mode-field diameter unstructured hollow fiber driven by the 1.5 µm signal beam delivering 2.25 mJ pulses, the phase matching cutoff extends well above the carbon K-edge, to ~400 eV. Using a 90 µm mode-field fiber diameter ARHCF – the smallest core commercially available 3 µm ARHCF – driven by the 3 µm idler beam delivering 0.775 mJ pulses, the signal from He is weak due to sub-optimal laser intensity: maximum HHG flux requires ionization of the gas to the critical ionization level for phase matching, as well as coherent buildup over several absorption lengths. However, using a 5x higher laser intensity (e.g. pulse energy of 4 mJ, or 2 mJ pulse with a 60 fs pulse duration), we predict much greater flux, peaking around 600 eV and extending to >1 keV, as has been observed previously for mid-IR HHG in unstructured waveguides [14]. In that work, a flux of ~$10^6$ photons/s/1% bandwidth at 1 keV (corresponding to ~$10^5$ photons/pulse/1% bandwidth) was measured from a ~10 mJ, 3.9 µm OPCPA. By scaling the repetition rate from 20 Hz to 1 kHz and accounting for the ~$\lambda^{-6}$ wavelength dependence of the single-atom yield, 3 µm driven HHG in a He-filled ARHCF will enable application-relevant flux of ~$10^9$ photons/sec/1% bandwidth up to keV photon energies.

As an exciting avenue towards higher 3 µm pulse energies, 2 µm wavelength lasers can be used to pump OPCPAs with higher pump-to-signal conversion efficiencies [42], making it possible to optimally drive He HHG. Alternatively, for the OPCPA system discussed here, leveraging the higher multi-mJ pulse energy and improved HHG conversion efficiency of the signal beam at 1.5 µm, SXR HHG with ~$10^9$ photons/s/1% bandwidth is achievable at photon energies around the carbon K-edge, as has already been demonstrated using Ti:sapphire driven OPA systems at similar wavelengths [17].

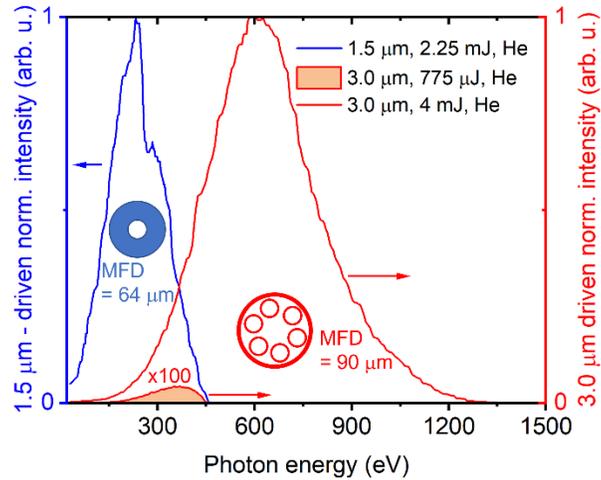

Fig. 6. Simulated helium HHG spectra for a 1.5 µm driver in a 64 µm MFD unstructured capillary (in blue) and for a 3 µm driver in a 90 µm ARHCF (in red). Note that the 3 µm 775 µJ data is magnified by one hundred times. While the spectra are normalized separately for 1.5 and 3 µm, the peak flux can be comparable if we use longer ARHCFs for 3 µm, since the decrease in single-atom yield from doubling the wavelength can be compensated for by the lower harmonic absorption at 600 eV compared to ~300 eV.

## 5. Conclusion & future work

In conclusion, we demonstrated a ytterbium-pumped optical parametric chirped pulse amplifier laser system that employs an all-fiber front-end and periodically-poled lithium niobate that operates near the pulse fluence limits of current commercially-available PPLN nonlinear media. The OPCPA delivers 3 µm wavelength pulses with 775 µJ energy at 1 kHz repetition rate, with transform-limited 120 fs pulse duration, diffraction-limited beam quality, and ultrahigh 0.33% rms energy stability over >18 hours. The long-term stability of this OPCPA system enabled the generation of ultrastable soft X-ray high harmonics in an anti-resonant hollow core fiber, backed with up to 10 bar of Ar or $N_2$, at photon energies up to the carbon K-edge (284 eV), with high beam quality and long-term operation over periods of months. Finally, we discussed routes for achieving the optimal theoretical soft X-ray high harmonic flux in a He-filled ARHCF, driven at higher laser intensities either by the signal beam at 1.5 µm, or by the idler at 3 µm.

## 6. Funding

We gratefully acknowledge support from the U.S. Department of Energy (DE-SC0020752) for the fiber front-end development, an NSF MRI Award 1828705 for the 3 µm beamline, and a Moore Foundation Award 10784 for the 1.5 µm beamline.

## 7. Acknowledgments

We acknowledge helpful discussions with Rodrigo Martin Hernandez for the development of the HHG-code and Tsung-Han Wu for assistance with early fiber splicing for the OPCPA front-end.

## 8. Disclosures

MMM: Kapteyn-Murnane Laboratories (I,S), HCK: Kapteyn-Murnane Laboratories (F,I,E,S).

## 9. Data availability

Data underlying the results presented in this paper are not publicly available at this time but may be obtained from the authors upon reasonable request.